%% file: paper.tex
\documentclass{elsart}
\usepackage{epsfig}
\usepackage{rotate}
\usepackage{amssymb}
\usepackage{amsmath}

\begin{document}
\begin{frontmatter}
\title{\Large \boldmath An apparatus to search for 
 mirror dark matter via the invisible decay of orthopositronium in vacuum.
%Design of a  high-efficiency pulsed positron
%  beam and an experimental apparatus to search for 
%the dark matter of a mirror-type.\\
%or\\
}
%\centerline{\bf (DRAFT, \today)}

%\centerline{\bf NOMAD Collaboration}
\author[Zuerich]             {A.~Badertscher}
\author[INR]                 {A.S.~Belov}
\author[Zuerich]             {P.~Crivelli}
\author[Zuerich]              {M.~Felcini}
\author[Zuerich]              {W.~Fetscher}
\author[INR]               {S.N.~Gninenko}
\author[INR]               {N.A.~Golubev}
\author[INR]               {M.M.~Kirsanov}
\author[IHEP]              {L.L. ~Kurchaninov}
%\author[LAPP]              {P.~N\'ed\'elec}
\author[LAPP]            {J.P.~Peigneux}
%\author[INR]               {V.~Postoev}
\author[Zuerich]    {A.~Rubbia }\footnote{contactperson, e-mail
 address: Andre.Rubbia\char 64cern.ch}
\author[LAPP]              {D.~Sillou}

\address[Zuerich]        {ETH Z\"urich, Z\"urich, Switzerland}
\address[INR]            {Institute for  Nuclear Research, INR Moscow, Russia}
\address[IHEP]           {Institute of High Energy Physics, Protvino, Russia}
\address[LAPP]           {CNRS-IN2P3, France}

%\clearpage
\begin{abstract}
Mirror matter is a possible dark matter candidate.
%Among several interesting motivations for experimenting with 
%orthopositronium ($o-Ps$) in vacuum the most exciting one is probably 
%related to the search for the dark matter of a  mirror-type. 
It is predicted to exist if parity is an unbroken symmetry of the vacuum.
The  existence of the mirror matter, which in addition to gravity
is coupled 
to our world through photon-mirror photon mixing, would result
in  orthopositronium ($o-Ps$) to mirror orthopositronium ($o-Ps'$)
 oscillations.  The experimental signature of this effect is the 
invisible decay of $o-Ps$ in vacuum.

This paper describes the design of the new
experiment for a  search for the $o-Ps\to invisible$ decay  
in vacuum with a sensitivity in the branching ratio of 
$Br(o-Ps\to invisible)\simeq 10^{-7}$, which is an order of 
magnitude better than the present limit on this decay mode 
from the Big Bang Nucleosynthesis. The experiment is based on
 a high-efficiency pulsed slow positron beam, which is also applicable
for other experiments with $o-Ps$, and (with some modifications) for 
applied studies. 
Details of the experimental design and of a new pulsing method,
 as well as 
preliminary results on  requirements for 
the pulsed beam components are presented.
The effects of $o-Ps$ collisions with the cavity walls as well as the influence
of external fields 
 on the $o-Ps\to o-Ps'$ oscillation probability are also discussed.

% design of a new high-efficiency pulsed slow positron beam for 
%xperiments with orthopositronium in vacuum is described.
%The new method is based on a RF waveform pulse applied to a double 
%gap buncher 
%system for adjusting the time-of-flight of positrons. Simulations show that 
%compression of positrons emitted from the moderator 
%during 300 ns  into a pulse of 2 ns width is achievable.
%Preliminary results on the design and requirements for 
%the pulsing system components are presented.   

\end{abstract}
\end{frontmatter}

\input{introduction.tex}

\input theory.tex

\input setup.tex

\input beam.tex

\input pulsebeam.tex

\input simulations.tex

\input requirements.tex

\input montecarlo.tex

\input sensitivity.tex

\input conclusion.tex

{\bf Acknowledgments}

We express our thanks to Z. Berezhiani,  R. Foot, S.L. Glashow,
N.V. Krasnikov, V.A. Matveev and R. Mohapatra for many stimulating
communications and discussions. 
The help and  collaboration  with N. Alberola (LMOPS), 
A. Gonidec, G. Roubout and J. Wolf from CERN, 
 C. Wyon (CEA), and L. Knecht on DC beam prototype construction
 are greatly appreciated. 
We would like to thank Ph. Sp$\ddot{a}$lig (PSI) for the preparation of the
 $^{22}$Na source, A. Skassirskaya (INR) for her help with the beam
simulations and D. Taqqu (PSI) for his 
interest and  useful discussions.
Support of the Swiss National Foundation, ETH Z\"urich and 
Institute for Nuclear Research, Moscow is  
gratefully acknowledged.

\input bibliography.tex
\end{document}

%% file: introduction.tex
\section{Introduction}

The dark matter problem provides one of the strongest indications for 
 physics beyond the Standard Model. 
Although the unknown physics is usually addressed  
in a direct manner in  high energy experiments,  
new results may also be
expected from precision experiments at lower energies.\\ 
Orthopositronium ($o-Ps$, the triplet $e^+e^-$-bound state), 
is a particularly interesting system for such an approach [1-3]. 
For example, it has been shown recently that experiments 
searching for $ invisible $ decays of $o-Ps$ with the
(currently achievable) level of sensitivity in the branching ratio 
$Br(o-Ps\to invisible) \simeq 10^{-8}-10^{-9}$
have significant  discovery potential \cite{gka}. 
An observation of $o-Ps\to invisible $ decay  would unambiguously 
signal new physics phenomenon which could be induced  either by the 
existence of extra dimensions \cite{gkr1}, or of
 fractionally charged particles [5-7], 
or of light gauge bosons \cite{gka}. Other interesting  
experiments with $o-Ps$ are motivated 
by tests of  high order QED corrections to the
 $o-Ps$  decay rate \cite{afs1}, searching for a violation of fundamental symmetries
 in positronium annihilation \cite{vetter}, 
tests of antimatter gravity in the
free gravitational
fall of positronium \cite{mills1}, the  possibility to observe 
positronium Bose-Einstein condensation \cite{mills2}
and others. Among them the most exciting one is probably related to the 
search for the dark matter of a mirror-type.

Mirror matter is predicted to exist if parity is an
unbroken symmetry of nature.
The idea was originally  discussed  by 
Lee and Yang \cite{ly} in 1956, who  suggested that the transformation 
in the particle space corresponding to the space inversion  ${\bf x\to -x}$
should not be the usual transformation  {\it  P}, but  {\it
  PR}, where  {\it  R} 
corresponds the transformation of a particle (proton\cite{ly}) into a
reflected state in the mirror particle space.  
After observation of parity nonconservation Landau assumed \cite{dau}
that  {\bf \it  R=C}, i.e. he suggested to identify antiparticles with the 
mirror matter, but then  {\bf \it CP} must be conserved which we know is not 
the case.
The idea was further developed 
by A. Salam \cite{salam}, and was clearly formulated in 1966
as a concept of the Mirror Universe 
by Kobzarev, Okun and Pomeranchuk \cite{kop}. 
In this paper it was shown that ordinary and mirror matter can 
communicate  predominantly through gravity and proposed that the 
mirror matter objects  can be present in our universe.   
  
Since that time the mirror matter concept has found many interesting 
applications and developments. In 1980 it has been boosted by 
superstring theories with $E_8\times E_8^{'}$ symmetry, where the particles 
and the symmetry of interactions in each of the $E_8$ groups are identical.
Hence, the idea of mirror matter can be naturally accomodated.
%One could describe the motivation for the mirror matter in the 
%following way \cite{venj}. The space inversion and any other 
%geometric symmetry from the Poincare'-group is represented not
%by a one operator {\bf \it  P}, but a whole class of operators  
%{\it  PR}, where {\it  R} forms and internal symmetry group of the system.
%The space of the particles is assumed to be a representation of 
%the extended Poincare' group, i.e. one which includes the space 
%coordinate inversion  ${\bf x \to -x}$. 
%Since the space inversion and time shifts commute, the corresponding 
%operations in the particle space,  {\it  PR} and Hamiltonian 
%{\it H},  must also commute.
%It implies that parity defined as eigenvalue of  {\it  PR}
% must be an integral of 
%motion for a closed system. It can be suggested while  {\it P}
% is the usual coordinate reflection, and  {\it  R} is the transformation 
%of an ordinary particle to the mirror one. Thus, parity 
%is conserved in the total space of ordinary and mirror particles.
Today's  mirror matter models exist in two basic versions. The symmetric 
version proposed early was further developed and put into a modern context 
by Foot, Lew and Volkas \cite{flv}. The asymmetric 
version was proposed by Berezhiani and Mohapatra \cite{bm}.
More detailed discussions of mirror matter models 
can be found in Ref.\cite{ber}.

In the symmetric mirror model the idea is 
that for each ordinary particle, such as the photon, electron, proton
and neutron, there is a corresponding mirror particle, of 
exactly the same mass as the ordinary particle. 
The  {\it PR} operator interchanges the ordinary particles with the
mirror particles so that the properties of the mirror
particles completely mirror those of the ordinary particles.
For example the mirror proton and mirror electron are stable and 
interact with the mirror photon in the same way in which the
ordinary proton and electron interacts with the ordinary photons.
The mirror particles are not produced
in laboratory experiments just because they couple very
weakly to the ordinary particles. In the modern language of gauge
theories, the mirror particles are all singlets under 
the standard $G \equiv SU(3)\otimes SU(2)_L \otimes U(1)_Y$
gauge interactions \cite{flv}. Instead the mirror
particles interact with a set of mirror gauge particles,
so that the gauge symmetry of the theory is doubled,
i.e. $G \otimes G$ (the ordinary particles are, of 
course, singlets under the mirror gauge symmetry)\cite{flv}.
Parity is conserved because the mirror particles experience
$V+A$ (i.e. right-handed) mirror weak interactions
while the ordinary particles experience the usual $V-A$ (i.e.
left-handed) weak interactions.  
%Ordinary and mirror
%particles interact with each other by
%gravity. At the present time there is some experimental evidence
%that mirror matter exists coming from cosmology as well as from
%the neutrino physics anomalies\cite{foot3}. 

It was realized some time ago by Glashow\cite{gl}, that
the orthopositronium system provides a sensitive
way to search for the mirror matter. 
Glashow's idea is that if a small kinetic mixing of the ordinary and mirror
photons exists \cite{bob}, it would mix ordinary and mirror 
orthopositronium, leading to maximal orthopositronium -
mirror orthopositronium oscillations, see Figure 1. Since mirror $o-Ps'$
decays predominantly into three mirror photons these oscillations 
result in $o-Ps\to invisible$ decay in vacuum.
Remember, that  due to the odd-parity under
{\it C} transformation  $o-Ps$  decays predominantly into three photons. 
 As compared to the singlet ($1^1S_0$) state (parapositronium),
 the small $o-Ps$ decay rate (due to the phase-space and  additional
$\alpha$ suppression factors) gives an enhancement factor $\simeq 10^3$,
making it more sensitive to an admixture of this 
new interaction \cite{gka,dobr,rich}.

%On the experimental side,
%there have been a number of measurements
%of the lifetime of orthopositronium. The most accurate
%measurements are given in the table below:
%\vskip 1.0cm

%{\begin{center}
%\begin{tabular}{|l|l|l|l|}
%\hline
%Reference$\;\;\;\;\;\;\;$
%&$\Gamma_{oPs} (\mu s^{-1})$ $\;\;\;$ 
%&Method$\;\;\;\;\;\;\;\;\;\;\;\;$
%&$\Gamma_{coll}$$\;\;\;\;\;\;\;\;\;\;\;\;$\\
%\hline
%Ann Arbor\cite{aa1}&$7.0482\pm 0.0016$&Vacuum Cavity&$\sim (3-10)\Gamma_{oPs}$\\
%Ann Arbor\cite{aa2}&$7.0514\pm 0.0014$&Gas&$\sim 10^3 \Gamma_{oPs}$\\
%Tokyo\cite{tok}&$7.0398\pm 0.0029$&Powder&$\sim 10^4 \Gamma_{oPs}$\\
%\hline
%\end{tabular}\end{center}}

%Table Caption:
%Some measurements of the orthopositronium lifetime.
%The last column is an estimate of the mean scattering
%length of the orthopositronium in the experiment.

%\vskip 1cm

Photon-mirror photon kinetic mixing
is described by the interaction Lagrangian density

\begin{equation}
L = \epsilon F^{\mu \nu} F'_{\mu \nu},
\label{ek}
\end{equation}
where $F^{\mu \nu}$ ($F'_{\mu \nu}$) is the field strength 
tensor for electromagnetism (mirror electromagnetism).
The effect of ordinary photon - mirror photon kinetic mixing
is to give the mirror charged particles a small electric
charge\cite{flv,gl,bob}. That is, they couple to ordinary photons with
charge $2\epsilon e$\footnote{Note that the direct experimental
bound on $\epsilon$ from searches for `milli-charged' particles
is $\epsilon \stackrel{<}{\sim} 10^{-5}$
\cite{davidson,prinz}.}.

\begin{figure}[htb]
\begin{center}
\hspace{-1.cm}{\epsfig{file=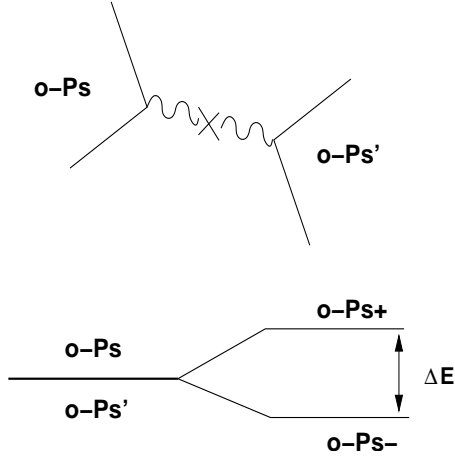,width=60mm,height=60mm}}
\end{center}
\vspace{1.0cm}
 \caption{\em The double degeneracy between orthopositronium mass eigenstates
of ordinary o-Ps and mirror o-Ps' is broken when a small mixing 
(upper picture) term is included. }
\label{mixing}
\end{figure}
Orthopositronium is connected via a one-photon
annihilation diagram to its mirror version ($o-Ps'$).
This breaks the degeneracy between $o-Ps$ and $o-Ps'$ so that
the vacuum energy eigenstates are $(o-Ps + o-Ps')/\sqrt{2}$ and 
$(o-Ps - o-Ps')/\sqrt{2}$,
which are split in energy by $\Delta E = 2h\epsilon f$, 
where $f = 8.7\times 10^4$ MHz is the contribution to the
ortho-para splitting from the one-photon annihilation diagram
involving $o-Ps$ \cite{gl}.
Thus, the interaction eigenstates are maximal combinations
of mass eigenstates which implies that $o-Ps$ oscillates
into $o-Ps'$ with probability: 
\begin{equation}
P({\rm o-Ps} \to {\rm o-Ps'}) = \sin^2 \omega t, 
\end{equation}
where $\omega = 2\pi\epsilon f$.

In the simplest case of $o-Ps \to o-Ps'$ 
oscillations in vacuum\cite{gl},
because the mirror decays are not detected,
this leads to an {\it apparent} increase in the decay
rate, since the number of $o-Ps$, $N$ satisfies
\begin{equation}
N = \cos^2 \omega t e^{-\Gamma_{sm} t}
\simeq exp [-t(\Gamma_{sm} + \omega^2t)],
\end{equation}
where
$\Gamma_{sm}=7.039934(10)\mu s^{-1}$ \cite{afs} is the Standard Model 
decay rate of $o-Ps$ (i.e. when
the oscillation length goes to infinity). 
Thus $\Gamma^{eff} \approx \Gamma_{sm}\bigl(1 + \omega^2/\Gamma_{sm}\bigr)$.

The above calculation is not applicable to
an experiment, where the positronium is confined in a cavity,
 because in this case
the collision rate is not zero and the 
loss of coherence due to the collisions must be included 
in the calculation of Eq.(3) \cite{gnin,fg}.
 Note,  that the probability 
$P({ o-Ps} \to { o-Ps'})$  can also be affected by an 
additional splitting of $o-Ps$ and $o-Ps'$ states 
by an external electric or magnetic  field \cite{gnin}.
This is quite similar to the phenomenon of 
$n-\overline{n}$  \cite{kuzj} or muonium to antimuonium 
oscillations \cite{fw} in various  environments,
see section 2 for a discussion of these effects.
% The effect that collisions damp the oscillations
%is well known and in the limit where the collision rate is much larger
%than the decay rate (or oscillation frequency; whichever is smaller)
%the effect of the oscillations becomes negligible (Quantum Zeno effect).
%Let us assume for now that the theory computation is accurate
%(i.e. the unknown quantity $B_0$ is not anomolously large).
% The agreement
%with the Tokyo experiment can be explained because of the
%very large collision rate of the orthopositronium in the powder.
%However because of the two different collision rates of the 
%two Ann Arbour experiments, they cannot both be explained.
%If we ignore this gas result then
%the discrepancy between the theory/Tokyo results and the
%Ann Arbour vacuum cavity experiment can be explained by
%the orthopositronium-mirror orthopositronium oscillation
%mechanism. 

Recently, Foot discussed implications of 
the DAMA \cite{dama} and CRESST \cite{cresst} experiments for mirror matter-type dark matter 
which is coupled to ordinary matter through the interaction of Eq.(1)
 \cite{foot1} (for references related to mirror matter see e.g. 
\cite{mirror}).  He has shown  that the
 annual modulation signal obtained by the DAMA/NaI experiment, 
as well as the CRESST data, can be
       explained by mirror matter-type dark matter if the photon-mirror photon
mixing strength is in the region 
\begin{equation} 
\epsilon  \simeq 4 \times 10^{-9}
\label{ft1}
\end{equation}

 Interestingly, this  value of $\epsilon$ is also 
consistent with all other known experimental and cosmological bounds 
including SN1987a \footnote{The SN1987a limit $\epsilon < 10^{-9.5}$ 
obtained in Ref.{\cite{kmt}} is actually much weaker. 
For a more detailed discussion of this constraint, see
  Ref.{\cite{fiv}}} and the standard Big Bang Nucleosynthesis (BBN) 
bound  \cite{cg}.  It is also in the  range of naturally small 
$\epsilon$-values motivated by grand unification models \cite{ber}. 

If $\epsilon$ is
as large as in Eq.(\ref{ft1}),
the branching ratio $ Br(o-Ps\to invisible)$ for invisible decays of 
orthopositronium in vacuum is of the
order (see Section 4): 
\begin{equation}
Br(o-Ps\to invisible)\simeq 2\times 10^{-7}
\label{ft2}
\end{equation}
For comparison,  the BBN limits \cite{cg} deduced from the successful 
prediction of the primordial $^4$He abundance are
\begin{equation}
\epsilon < 3\times 10^{-8}
\label{bbn1}
\end{equation}
and 
\begin{equation}
Br(o-Ps\to invisible) < 10^{-5}
\label{bbn2}
\end{equation}
respectively.

%If the photon - mirror photon mixing is as
%large as suggested 
%here then there will be a number of interesting implications.
%For example, any mirror matter in the center of the sun can become quite
%hot due to the absorption of ordinary photons.
%The subsequent radiation of mirror photons by the mirror matter
%would be absorbed by the surrounding ordinary matter providing
%a new source of energy transport in the solar interior.
%Another implication of the large photon - mirror photon kinetic
%mixing is that it may be large enough (depending on
%the chemical composition of the mirror planet) to make mirror planets
%opaque to ordinary photons.
%While it is fun to speculate about the effects of large
%photon - mirror photon kinetic mixing, real progress
%requires experiments.\

% Concerning the
% mixing strength $\epsilon$ an experimental 
%limit can be derived from some of earlier papers \cite{mitsui}.\

The first experiment on the $o-Ps \to invisible$ decay, motivated by a puzzle in the $o-Ps$
decay rate (see below), was performed  
a long time ago \cite{atojan}, and then repeated 
with higher sensitivity  \cite{mits}.\ The results exclude
 contributions to the $o-Ps$ decay rate from 
invisible decay modes (such as $o-Ps\rightarrow \nu \nu$, millicharged 
particles, etc..) at the level of $Br(o-Ps\rightarrow 
invisible) <3\cdot 10^{-6}$, 
but are not very sensitive to the $o-Ps \to o-Ps'$ oscillation
mechanism because of the high collision rate in these experiments. 
Indeed the limit on $\epsilon$  extracted from the results of 
ref. \cite{mits}, taking into account the 
suppression collision factor, is $\epsilon < 10^{-6}$ \cite{gnin} and is not 
strong enough to exclude a possible mirror-matter contribution at the level 
comparable with the BBN limit of Eq.(\ref{bbn1}).\\
% Thus, an experiment with significantly higher sensitivity would
% be interesting  to search for the mirror matter effect.\ 
 Given the indications for the mirror world
coming from dark matter\cite{foot1} and the neutrino physics 
\cite{venj,foot3}, as well as the intuitive expectation
that nature could  be left-right symmetric, 
it is obviously important to  determine experimentally
whether orthopositronium is a window on the mirror world. 
%Thus, if it were experimentally proven that $\epsilon$ is
%as large as Eq.(\ref{ep}) then it would presumably mean that
%some of the assumptions of big bang nucleosynthesis would
%have to be modified.
%We hope that future experiments will shed light on this issue \cite{crivelli}.
Since there are no firm predictions for  $\epsilon$,
experimental searches for $o-Ps\to invisible$ have to be  performed  
with a  sensitivity as high as possible.\ In particular, the  question
whether  the sensitivity for the interesting branching ratio of Eq.(\ref{ft2})
is experimentally reachable has to be studied.\\
 The goal of this work is to present a design of the
 experiment for a search for $o-Ps\to invisible$ decay in vacuum  
with a sensitivity better than the corresponding BBN limit
$Br(o-Ps\to invisible) < 10^{-5}$
and high enough to check the prediction of Ref.{\cite{foot1}},
$Br(o-Ps\to invisible) \simeq 2\times 10^{-7}$.
The experiment requires the production  and subsequent 
decays of $o-Ps$'s to occur in vacuum \cite{gnin,fg}, and hence, the
 use  of a specially  designed  slow positron beam 
operating preferably in a pulsed mode to enhance the signal-to-noise ratio
for the efficient tagging of $o-Ps$ production \cite{gninops}.
The experimental signature of $o-Ps\to invisible$ decay
is the absence  of an energy deposition 
which is expected from the ordinary o-Ps annihilation
in a 4$\pi$ calorimeter
surrounding the positronium formation target.\
We show that this signature 
 is clean and that the signal events can be identified with a high 
confidence level due to the efficient tagging of the positron
 appearance in the 
target and the high-efficiency   measurement of its  annihilation energy.\
It should be noted that recent substantial efforts devoted to the
theoretical and experimental determination of the QED $o-Ps$ properties,  
hopefully results in a solution of the long-standing  discrepancy between the 
measured  and predicted orthopositronium decay rate in  vacuum, 
see e.g. \cite{ds,asai2,nico}. However, the 
current level of the theoretical precision achieved by Adkins et al.
 \cite{afs1,afs}
is about  two orders of magnitude better than the 
experimental one \cite{asai2,nico,vzg,asai1}. Thus,   further 
positron beam based  experiments to measure the
$o-Ps$ decay rate in vacuum are required and are of great  interest
to test high-order QED corrections.\\
The paper is organized as follows. 
In section 2 we consider the $o-Ps\to o-Ps'$  oscillations and effects of 
various environments on it.
In section 3 we report on 
the design of the experimental setup to search for the
$o-Ps\to invisible $ decay in vacuum.\ 
The  description of the detector components including the design of a 
high-efficiency pulsed positron beam is 
presented in sections 3 and 4. The  preliminary simulation
results and  expected sensitivity level 
 are discussed in sections 5 and 6, respectively.   
section 7 contains concluding remarks.

%% file: theory.tex
\section{$o-Ps\to invisible$ decay rate in a vacuum cavity}

In the simplest case of $o-Ps \to o-Ps'$ 
oscillations in vacuum \cite{gl} the branching ratio occurring during a long
enough observation time can be calculated as   

\begin{equation}
Br(o-Ps\to invisible) = \frac{2(2\pi \epsilon f)^2}{\Gamma^2_{sm} + 
4(2\pi \epsilon f)^2}
\label{br}
\end{equation}

% In this case,
%because the mirror decays are not detected,
%this leads to an {\it apparent} increase in the decay
%rate, since the number of o-Ps, $N$ satisfies
%\begin{equation}
%N = \cos^2 \omega t e^{-\Gamma_{sm} t}
%\simeq exp [-t(\Gamma_{sm} + \omega^2t)],
%\end{equation}
%where
%$\Gamma^{sm}$ is the standard model decay rate of o-Ps (i.e. when
%the oscillation length goes to infinity).
%Thus $\Gamma^{eff} \approx \Gamma^{sm} + w^2/\Gamma^{sm}$.

 Eq.(\ref{br}) may not be applicable to
an experiment in a cavity.
It is well known that collisions damp the oscillations, e.g. 
 in the limit where the collision rate is much larger
than the decay rate (or oscillation frequency, whichever is smaller)
the effect of the oscillations becomes negligible.
In addition, external fields might result in a
loss of coherence due to additional splitting of mass eigenstates. Thus, 
 their effect must be included \cite{gnin,fg}.\

% The agreement of Michigan result with the Tokyo experiment can be 
%used to set a limit on photon-mirror photon mixing strength, similar to 
%work \cite{gnin}.
% The agreement of Michigan result with the Tokyo experiment can be 
%explained because of the
%very large collision rate of the orthopositronium in the powder.
%Because of the two different collision rates in these 
%two  experiments, they cannot both be explained.

% ground state of orthopositronium (o-Ps) decays predominately into
%3 photons with a theoretical decay rate computed to be
%(see e.g. \cite{ck} for a review)
%\begin{eqnarray}
%\Gamma  &=&
%{2(\pi^2 - 9)\alpha^6 m_e \frac 9\pi}\big[
%1 - 10.28661{\alpha\over \pi} - {\alpha^2 \over 3}ln{1 \over \alpha}
%+ B_0 \left({\alpha \over \pi}\right)^2
%- {3\alpha^3 \over 2\pi}ln^2 {1\over \alpha}+ \nonumber \\
%&+&O(\alpha^3 ln\alpha)\big]
%\simeq (7.0382 + 3.9\times 10^{-5}B_0)\mu s^{-1}.
%\label{xxx}
%\end{eqnarray}

%The $B_0$ term
%parametrizes the non-logarithmic two-loop effects which have been 
%recently calculated \cite{afs}.

%The agreement  between the vacuum Michigan and Tokyo results 
% can be interpreted to extract limit on 
%the orthopositronium-mirror orthopositronium oscillation
%mechanism. 
Let us  first 
 consider the case where the collision rate is much larger than the
 decay rate,
$\Gamma_{coll} \gg \Gamma_{sm}$ \cite{fg}, then the evolution
of the number of orthopositronium states, $N$, satisfies:
\begin{equation}
{dN \over dt} \simeq
- \Gamma_{sm} N - \Gamma_{coll}N\rho,
\label{1}
\end{equation}
where 
the second term is the rate at which o-Ps oscillates into $o-Ps'$
(whose subsequent decays are not detected). In this term,
$\rho$ denotes the average oscillation probability
over the collision
time. That is,
\begin{equation}
\rho
\equiv \Gamma_{coll}\int^t_{0} e^{-\Gamma_{coll}t'} \sin^2 \omega t' dt' \simeq
\Gamma_{coll}\int^t_0 e^{-\Gamma_{coll}t'} (\omega t')^2 dt',
\end{equation}
where we have used the constraint
that the oscillation probability is small, i.e. $\omega t \ll 1$.
%(which must be the case given that the discrepancy between
%say the vacuum cavity experiment and Tokyo experiment is less than 
%1 percent).
As long as $t \gg 1/\Gamma_{coll}$, a reasonable approximation
 for the vacuum cavity experiment,  then
\begin{equation}
\rho \simeq {2\omega^2\over \Gamma_{coll}^2 }.
\end{equation} 
Thus, substituting the above equation into Eq.(\ref{1})
we have
\begin{equation}
\Gamma^{eff} \simeq \Gamma_{sm} + {2\omega^2 \over \Gamma_{coll}}
= \Gamma_{sm}\left(1 + {2\omega^2 \over \Gamma_{coll}\Gamma_{sm}}\right). 
\end{equation}
The difference between the higher decay rate measured in the vacuum
cavity experiment,
relative to the value predicted by theory,  can be expressed as
\begin{equation}
\Gamma_{exp}- \Gamma_{sm} \simeq {2\omega^2 \over \Gamma_{coll}\Gamma_{sm}} 
\end{equation} 
For the cavity size and the $o-Ps$ emission spectrum in a recent experiment on 
the $o-Ps$ decay rate in vacuum \cite{vzg}, we estimate that  
 $\Gamma_{coll} \lesssim 3\Gamma_{oPs}$,
which, neglecting the contribution from external fields 
(which are in fact negligible in this case\cite{gnin}),
implies that
\begin{equation}
\omega^2 \sim 2\times 10^{-3}\Gamma^2_{oPs} \Rightarrow
\epsilon \lesssim 10^{-6}. 
\label{ep}
\end{equation}
Thus, the limit of Eq.(\ref{ep}) is still not strong enough compared to 
the BBN one of Eq.(6).

%The much larger collision rates, e.g.
%of the Tokyo (and gas) experiments,  means that the oscillations can have no 
%effect on these experiments. Thus the value of the vacuum
%experiment relative to the 
%theory/Tokyo results should be higher. 

In the presence of a static external electromagnetic field
the $o-Ps - o-Ps'$ degeneracy is broken and the probability that positronium 
decays as mirror $o-Ps'$, rather than ordinary $o-Ps$, is 
\begin{equation}
Br(o-Ps\to invisible)\simeq \frac{2(2\pi \epsilon f)^2}{\Gamma^2_{sm} + 
4(2\pi \epsilon f)^2+ \Delta^2}  
\end{equation}

\noindent where $\Delta$ represents an additional 
 oscillations damping factor combining
effects of collisions of $o-Ps$'s with the cavity walls, scattering on  
residual gas atoms and the influence of  external fields \cite{gkr2}.

Estimates obtained taking into account Zeemann and Stark effects 
in positronium and  $o-Ps$
 scattering in the Van der Waals potential of residual gas moleculas 
in the cavity show 
that   for the external magnetic field  $\simeq$ 100 G, the
electric field $\simeq 100$ V and  residual vacuum pressure  
$\lesssim 10^{-8}$ Torr,  result in  a value  
$\Delta < \Gamma_{sm}$ \cite{gkr2}. The most crucial parameter 
is the number $N$ of $o-Ps$ collisions during its lifetime 
in the cavity, which gives a suppresion factor $\simeq1/N$. Hence 
$N$ should be as small as possible, i.e. $N\leq 1$.

%% file: setup.tex
\section{Experimental setup to search for  the $o-Ps\to invisible$ decay}

\begin{figure}[htb!]
%\begin{center}
\vspace{-0cm}\hspace{-0.cm}{\epsfig{file=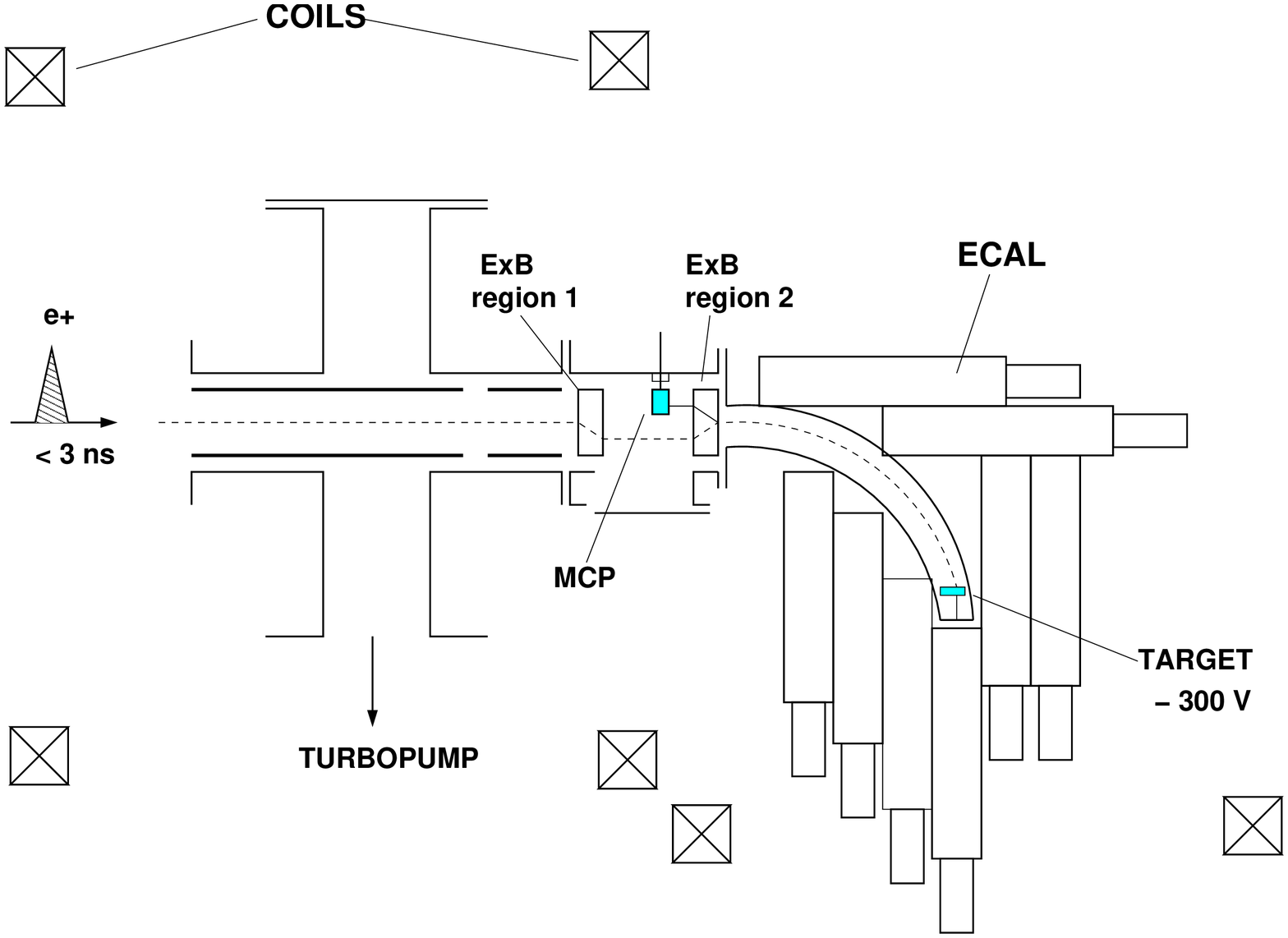,width=140mm,height=100mm}}
%\end{center}
\caption{\em Schematic diagram of the experimental setup.}
\label{tag}
\end{figure}

The experiment is designed with the goal to observe 
the  $o-Ps\to invisible$ decays, if its branching ratio
is greater than $10^{-7}$. 
Figure \ref{tag} shows a schematic view of the experimental setup in which 
 positrons from a pulsed beam \cite{gninops} 
are stopped in  the MgO target and either form positronium, i.e. $o-Ps$ or
$p-Ps$, or annihilate promptly into $2\gamma$'s. 
The secondary electrons (SE) 
produced by the positrons hitting the target are accelerated
 by the voltage applied to the target relative the grounded transport
 tube.  Then they are transported by a magnetic field in the
backward direction relative to the positrons 
moving in spirals along the magnetic field lines 
 and deflected to a microchannel plate (MCP)
 by a $E\times B$ filter as is shown in Figure \ref{traj}.

The trigger for  data acquisition 
is generated by a coincidence 
within $\pm$3 ns of a pulse from the MCP 
 and the signal from the 
pulsed beam, which is synchronized with the positron 
arrival time at the target. 

Accordingly, the apparatus
is designed with several distinct and separated parts:
i) a pulsed slow positron beam and a low-mass  target 
for efficient orthopositronium production in a vacuum cavity, 
ii) a  positron appearance tagging 
system with a high signal-to-noise ratio  based on a high performance MCP,
iii) an almost 4$\pi$  BGO crystal calorimeter (ECAL) surrounding the 
vacuum cavity for efficient detection of  annihilation photons. The 
 cavity has  as little wall mass as possible to minimize photon energy 
absorption.

The occurrence of the 
$o-Ps\to o-Ps' \to invisible$ conversion would appear as an excess
of events with energy deposition comparable with zero  in the 
 calorimeter above those expected  from the
prediction of the background.
In case of a signal observation 
the number of excess events could be cross-checked by 
 small variations of experimental conditions  which affect the 
$o-Ps\to o-Ps'$ transition rate but do not result in a
 loss of energy from ordinary positron annihilations. 
The identification of signal events 
 relies on a high-efficiency measurement of the energy deposition from the 
annihilation of positrons. 
Compared to the previous experiments on $o-Ps \to invisible$ decay
, the use of the BGO
calorimeter gives a significant reduction in total required ECAL mass
(see section 3.3).

To achieve a sensitivity in the branching ratio of $10^{-7}$ 
in a reasonable amount of data-taking time, the rate of 
$o-Ps$ decays per second
 has to be as high as possible consistent with minimal reduction of the
  $o-Ps\to invisible$ signal efficiency and acceptably 
small dead time. For the  pulsed positron beam
design presented  in Section 4,
 the trigger rate in the photon detector is expected to be 
$\simeq$ 100 Hz which is low enough to allow these 
events to be recorded without losses.

\subsection{Positron tagging system}

\begin{figure}[htb]
%\begin{center}
\vspace{-0cm}\hspace{1.cm}{  \epsfig{file=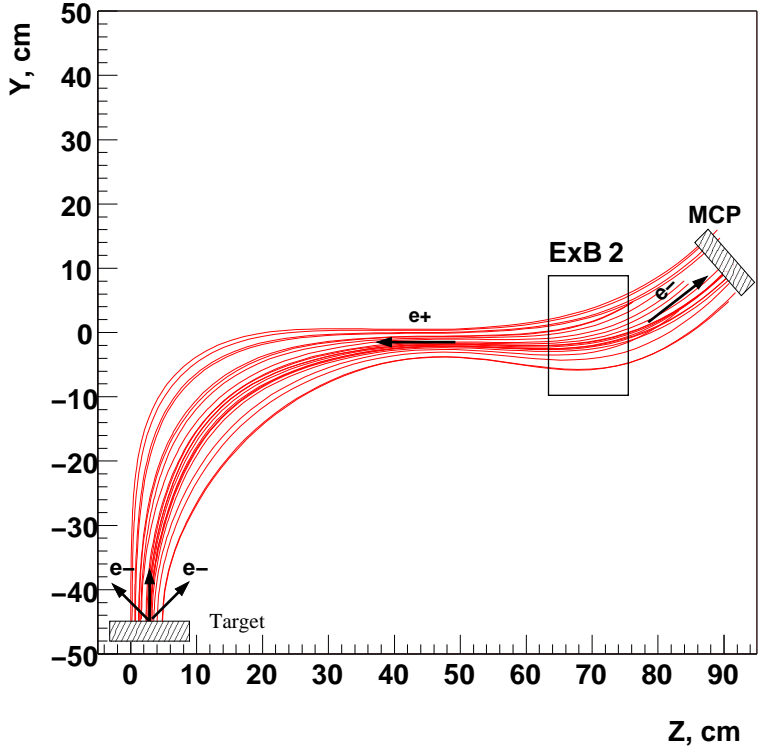,width=100mm,height=100mm,angle=-90}}
%\end{center}
\caption{\em Calculated trajectories  of secondary electrons in 
 the tagging system in Y-Z plane. The primary positron beam is
 increased in size for illustration purpose.}
\label{traj}
\end{figure}

The SEs produced by positrons hitting the $o-Ps$ production
target are used  to tag  the time of positron appearance in the target. 
 The positron tagging system is based on a high performance  
MCP  as a SE detector. 
The low energy SE emerging from the target are accelerated by an electric 
field and deflected to the MCP by the $E\times B$ filter, as shown in 
Figure \ref{tag}. The system works in detail  as follows. The
 pulsed beam of positrons with energy $\simeq$ 500 eV is guided by a 
magnetic field with a value of $B\simeq 100$ G and passes through a 
region with crossed 
electric and magnetic fields ($E\times B$ region 1 in the schematic
diagram of Figure \ref{tag}). 
The transversal electric field value is $E\simeq 500$ V/cm. 
%The {\em E} - field is created by a pair of plates with
%potentials $\pm$ 500 V applied to  each plate, respectively.   
Positrons  drift in the crossed electric and magnetic fields with a
 velocity given by
\begin{equation}
V_d = E\times B/ B^2
\end{equation}
 For the given values of the 
electric and magnetic fields the drift velocity is  $V_d = 7\times 10^3$ m/s
resulting in  the positron displacement about of 11 mm in the drift region
1. 
%Slow ions and electrons which are
%formed due to ionization of residual gas by charged particles in the
%vacuum system are separated from positrons in the $E\times B$ region 1. 
The electric field in region 2  has the 
same value but the opposite direction  relative to region 1.
%, the drift velocity direction of positrons is also reversed. 
As a result, downstream of the region 2 
positrons will move back to the axis if the transport system, separated
from  slow electrons and ions.
 Then, positrons are transported to the target in the curved magnetic field 
created by the coils. Preliminary results illustrating the method are shown
in Figure \ref{traj}, where the calculated trajectories of secondary 
electrons passing trough the $E\times B$ filter are shown
in the Y-Z plane for a wide primary positron beam. The 
energy spectrum of secondaries  is taken from Ref.\cite{se}, the angular 
distribution is assumed to be isotropic.   

 The target is a disk with a 
diameter of $\simeq$ 10 mm and a thickness of 0.1 mm. 
The SE acceleration  potential of -300 V is applied to the target. 
The target surface is coated by MgO for an efficient production of 
orthopositronium and an efficient secondary electron emission \cite{se}. 
%for incident positron energies of $\simeq$ 800 eV \cite{se}.  
The secondary electrons are transported  in the 
backward direction relative to the positrons, see Figure \ref{traj}, 
moving in spirals along the magnetic field lines. 
%In the $E\times B$ region 2 the electrons drift with the same velocity 
%as the positrons because the drift velocity is  independent 
%of the sign of the particle charge. In the region with 
%crossed field 2 the 300 eV electrons  will be displaced 14 mm 
%in transversal direction . 
Thus, trajectories of the secondary electrons will be spatially 
separated from the positron trajectories by a distance  estimated to  be 
$\simeq25$ mm, significantly larger then the 
diameter of the positron and the test electron beams ($\simeq$ 5 mm). The
MCP installed between  region 1 and region 2 (see Figure 2) detects the 
electrons. The background count rate of the MCP can limit the efficiency of the 
positron tagging. This and other sources of the background can be
suppressed by an appropriate  
choice of the MCP type and by using  a pulsed positron beam with a 
low duty cycle. 
%The operation in the pulsed mode 
% will also reduce other  sources of 
% background,  such as  cosmic rays, environmental radiation etc.  
The energy spectrum and the yield of 
SE are important for the tagging efficiency.
The SE emission is a surface effect, involving only a very 
thin layer of the target material in the process. Thus, the SE yield and the 
MCP output signal are proportional to the energy loss {\em dE/dx} of positron. 
The SE are emitted with an energy  up to 100 eV and a large angular 
spread. This results in a dispersion of their arrival time at the MCP. 
To reduce this effect the  
SE should be  accelerated immediately after 
production  by the electric field at the 
biased target.  Note that the MCP pulse shape could be 
used for additional  discrimination between the signals from positrons and 
the noise pulses.

The factors affecting the performance of the positron tagging system are
\begin{itemize}
\item the coefficient of secondary electron emission;
\item  the efficiency of the electron transport from the target to the 
  MCP, and the efficiency of the MCP itself;
\item the MCP noise level and the environmental background; 
\item  physical backgrounds, from e.g. beam interactions with residual gas
and cavity walls, with material of the $E\times B$ filters etc. 
accompanied by electron or ion production.
\end{itemize}
We have selected the MCP  Hamamatsu F4655-12 for the tagging system
because  it provides the best signal-to-noise ratio.
The  MCP signal rise time  is 
$\simeq$ 300 ps. The signal amplitude varies typically from 0.1 to 1 V
depending on the applied voltage, the MCP gain and the 
number of the emitted SEs.
The collection voltage between the MCP bottom surface and the anode
is not critical, a setting above 100 V is sufficient for $>95\%$
collection efficiency.

\subsection{$o-Ps$ production target}

A MgO coated film is planned to be used as $o-Ps$ production target.
The production rate of $o-Ps$ per single positron is about 25 \% at the 
positron energy of $\simeq$ 700 eV \cite{nico}. The MgO is also reported to have the
high secondary electron emission coefficient ( $\gtrsim 4$ ) at this 
energy. Another possibility is a high porosity and 
low density SiO$_{2}$ target \cite{vzg}.

To ensure a minimal probability of total absorption of annihilation photons, 
 the target itself and the surrounding components of the target 
region must be carefully 
designed. To minimize the thickness of dead material, in particular
the thickness of the beam pipe in this region is very important.
Based on simulation results a 1 mm thick Al pipe seems  the best
 choice ( see section 5).
One possible way to avoid the problem of dead material 
is to drill a hole in the central 
crystal (Figure \ref{tag}) and glue the beam pipe into it in 
 such a way that the $o-Ps$ production and decay region is surrounded 
almost completely by the crystal active medium . 
%Another solution is to put the calorimeter
%in a low vacuum and to make the beam pipe with as thin as possible
%walls ($\lesssim 1~mm$). 
%The design of this region is presently under study.  

\subsection{Photon detector}

%Note, that actually  the $\gamma$-detector serves to veto effectively the 
%positron annihilations into photons. It has been shown that 
%its inefficiency due to photoelectron statistics is 
%less then $10^{-9}$ for the energy threshold of $\simeq 50 $ keV 
%\cite{pc}.

The optimal choice of the $\gamma$-detector (ECAL) can be made by the following
 considerations. The total ECAL mass W is given roughly by 
\begin{equation}
W \simeq 4\pi/3 \rho L^3
\end{equation}

where $\rho$ and $L$ are respectively mass density and the radius of the ECAL
detector. We chose $L\simeq 20\lambda_{511}$, where $\lambda_{511}$
is the attenuation length of 511 keV $\gamma$'s. The relevant
parameters for 
different types of materials used in ECAL's are listed below in Table 1.

{\small
\begin{table}[h]
\begin{center}
\caption{Comparison between different types of ECAL.}
\vskip0.3cm
\begin{tabular}{|c|c|c|c|c|}
\hline
 ECAL type   & BGO&NaI&CsI(Tl)&Scintillator \\
&&&&plastic/liquid\\ 
\hline
$\lambda_{511}$ \hphantom{00}&  $\simeq$1 cm\hphantom{00} & $\simeq$2.5 cm\hphantom{00} &  $\simeq$1.9 cm\hphantom{00} & $\simeq$10 cm  \\
density,  $g/cm^3$  \hphantom{00}     & 7.1\hphantom{00} & 3.6\hphantom{00}&4.5\hphantom{00}&   1.0 \\
 ECAL mass, kg  \hphantom{00}   & $\simeq$240\hphantom{00}&  $\simeq$189
0\hphantom{00}& $\simeq$1034\hphantom{00}& $\simeq$33510\\
 $N_{\gamma}/511$ keV\hphantom{00} & $\simeq 4\cdot10^3$\hphantom{00}
 &$\simeq 20\cdot 10^3$\hphantom{00}
& $\simeq 10\cdot 10^3$\hphantom{00} & $\simeq 2\cdot10^3$\\
Hygroscopic \hphantom{00}                  & no\hphantom{00} & yes\hphantom{00} 
& \hphantom{00}slightly\hphantom{00}& no  \\
\hline
\end{tabular}
\end{center}
\end{table}
}
The required mass is minimal for a BGO ECAL due to its high effective
$Z$ (remember, the photo-absorption cross-section $\sigma\sim Z^5$). 
Another important feature of BGO's is that they are not hygroscopic,
thus, no additional dead material has to be introduced.

The schematic drawing of the $\gamma$-detector is shown in Figure \ref{tag}.
The $\gamma$-quanta produced in positron or positronium
 annihilation are detected 
by a (almost) 4$\pi$ BGO crystal calorimeter \cite{bgo}.
We plan to use BGO crystals also for measuring 
of the photon  time 
with respect to the arrival time, $t_0$  of the positron bunch  on the target.
The detector system consists of about 100 BGO crystal surrounding 
the vacuum beam pipe as  shown in Figure \ref{tag}.
%Each crystal has a 
%hexagonal shape with a diameter of 52 mm and a length of 200 mm.
%The internal ring of BGO counters
%is also used for the measurement of delayed annihilation of positrons 
%with respect to $t_0$. 
The full system is calibrated and monitored internally 
using the 511 keV annihilation line.  
For the crystal wrapped in aluminized mylar the light yield was
measured to be 
200$\pm$14 photoelectrons/1 MeV. This results in a probability of zero
energy 
detection due to Poisson fluctuation of the number of photoelectrons,
to be less 
than $10^{-11}$ for the zero energy signal defined as events with
energy deposition less than 50 keV \cite{pc,bader}.
This result justifies the selection of the BGO as the $\gamma$--detector.
The crystals, which have been lent to us by the Paul Scherrer
Institute 
(Villigen, Switzerland), have a hexagonal shape with a length of 20 cm
and an outer diameter of 5.5 cm, their original wrapping is a 0.75 mm
thick teflon. In order to reduce this amount of dead material, 
the inner ring of BGO's has been wrapped in 
a 2 $\mu$m thick foil, aluminized from both side with 1000 $\AA$ thick
layers. The required number of crystals, determined
with the simulations, 
provides an almost isotropically uniform thickness of 20--22 cm of BGO.

%% file: beam.tex
\section{High-efficiency pulsed positron beam}

In this section we report the preliminary design of a
high-efficiency  pulsed
slow positron beam for particle physics experiments with 
orthopositronium in vacuum.\ 
Our primary consideration is that the system should be of the magnetic 
transport type because this provides 
the simplest way to transport a slow 
positron beam  from the positron source to its  
target \cite{orli}. An electrostatic beam
 is also considered, but presently  the problems with 
construction, time schedule and increased cost dominate over the benefits. 
The basic working principals involved in the design of magnetic or 
electrostatic variable-energy DC positron beams are well known \cite{schultz}. 
The advantages and disadvantages of a beam formed and transported by a magnetic
field in comparison to the electrostatic one are also  known, see 
e.g. \cite{brusa}.\ 
%The advantage is that a high transport efficiency of the 
%positrons can be obtained, the disadvantage is that the geometrical 
%characteristics of the beam can not be fully controlled.  

Various techniques to create pulsed positron beams have been reported
with the main focus so far on material science applications \cite{col}.
Those attained by 
the Munich \cite{munich} and the Tsukuba \cite{tsuk} groups 
use RF power in the pulsed beam formation. However, acceleration by the sine
function of the RF electric fields is by no means the  optimal choice. 
The system based on this method requires a wide time window 
of chopping and accordingly the beam efficiency becomes small.

A pulsing system with a higher performance has  recently been proposed by 
Oshima et al.\cite{oshima}. 
The  main idea is the same as for the RF method: to adjust 
the time-of-flight for each positron according to the time it arrives 
at the accelerating point.  However, instead of applying a sinusoidal RF field,
a more suitable pulse shape of the electric field is generated such as an
approximate inverse parabol
function of time \cite{oshima}. This method has been further 
developed by Iijima et al \cite{iijima} for the material measurements in which 
the lifetime or time-of-flight of  orthopositronium atoms is close to
the $o-Ps$ lifetime in vacuum 
$\simeq$ 142 ns. For these applications it is necessary to  modify 
the originally proposed technique \cite{oshima} in order to generate 
higher intensity positron beams by accumulating positrons over a wider 
time span even though the bunch width becomes larger, but still 
much less than the typical measured timing intervals of $\simeq$ 100 ns. 
By using a high permeability buncher core a bunch width of 2.2 ns (FWHM) 
for 50 ns collection time and a repetition period of 960 ns 
has been achieved \cite{iijima}. 
The main problem encountered in this technique is the 
limitation of the voltage supplied by a post-amplifier to the buncher.

Our new pulsing method allows  to compress an initial positron pulse of 
300 ns to about 2 ns pulse width. The method
 relies on a positron velocity modulation by using a new   
double gap buncher technique.  The  detailed 
description of the system  components as well as their requirements 
are  presented.\

%% file: pulsebeam.tex
\subsection{Design criteria and overall system design}

The beam is supposed to be used for several different experiments 
with $o-Ps$ in vacuum. Thus, the final beam construction
 should  compromise several design goals which are crutial for them
 and which are summarized as follows:

\begin{itemize}

\item simple and not very costly  experimental apparatus, 

\item beam energy range from 100 eV to 1000 eV,

\item beam intensity of $ \simeq 10^4-10^5$ positrons per second,  

\item pulse duration at the target $\delta t_T \lesssim 3$ ns for an 
 initial pulse duration at the moderator $\delta t_M \simeq 300-400$ ns,

\item repetition rate 0.3-1.0 MHz,

\item high peak/noise ratio, (single) Gaussian shape of the pulse, 

\item beam spot size at the target  position is of the order of a few 
millimeters assuming 3-5 mm $^{22}$Na source diameter,

\item  minimal pumping time of the vacuum system,

%\item full automatic, computer controlled system. 

\end{itemize}

\begin{figure}[htb!]
\begin{center}
\hspace{-0.cm}{\epsfig{file=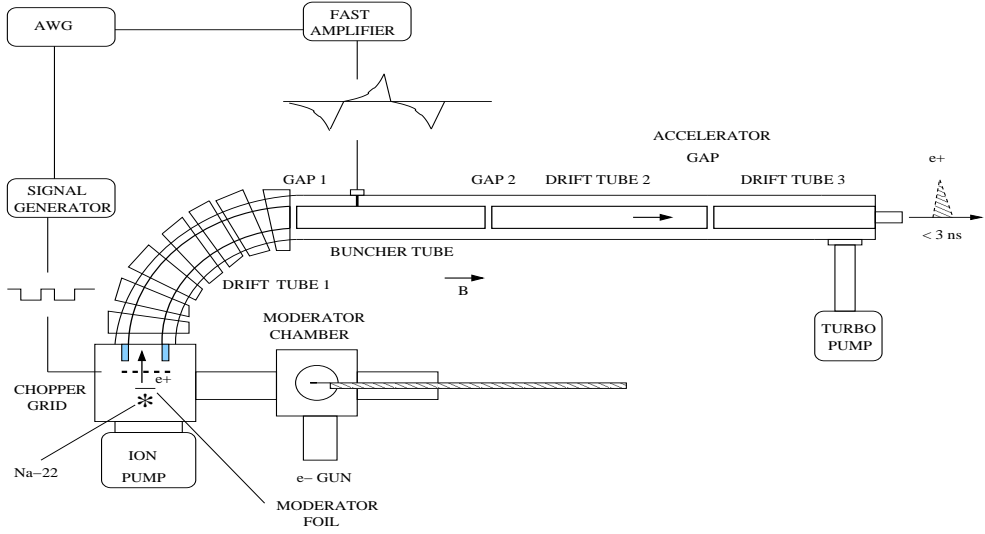,width=130mm,height=70mm,angle=0}}
\end{center}
\vspace{1.0cm}
 \caption{\em Schematic illustration of the magnetically transported pulsed  
positron beam.}
\label{beam}
\end{figure}

Figure \ref{beam} shows schematic  illustration of the pulsed
positron beam design.
The positron pulsing section consists of a
chopper and a buncher and is  based on positron
velocity modulation combined with the RF bunching technique.
A positive potential ($\simeq 100$ eV) is applied to the moderator foil
 in order to insure the proper energy of the positrons at the buncher input.
Initial positron pulses with duration 300 ns are formed with the chopper
grid placed 2 mm apart from  the moderator foil. The pulsed voltage with 
an amplitude about of +5V applied to the chopper grid relative the moderator
foil will  stop slow positrons with energy  about 3 eV emitted from the
moderator. Fast positrons emitted from the source are
eliminated from the beam by the velocity analyzer (90 degrees curved
solenoid, placed downstream the chopper). When the voltage applied to the
chopper grid is zero, the positrons come through the chopper grid and
are accelerated in the gap between the chopper grid and first drift
tube (see Figure \ref{beam}).  Thus, positron pulses with a duration of 300 ns are
produced by this way. The chopper voltage pulses can be produced by a
standard fast-signal generator.

In the gap between the drift tube 1 and the buncher tube the
velocity of positrons from the 300 ns pulse is modulated by a nonlinear pulsed
voltage applied to the buncher tube relative to the drift tubes.
The buncher tube length is determined by a distance-of-flight of 
 positrons entering the buncher during 300 ns. In a second gap
between the buncher tube and a drift tube 2 the positron velocity is 
modulated  again by the same voltage pulse applied to the buncher.

The buncher voltage pulse is produced by an  arbitrary waveform generator
(AWG) connected to a fast post-amplifier. The pulse shape for the
two-gap buncher is  determined by calculations described in Section 3. 
In accordance with Liouville's theorem the compression ratio in this case
is determined by the ratio of the final and initial energy dispersion in the
positron beam pulse. Experimentally measured initial energy dispersion of
the moderated positrons is about 2 eV (FW at 10\% of maximum) \cite{chen}. 
Taking into account that the final
energy spread in the given two gap buncher is about 200 eV we get 
the expected compression ratio of $\simeq$ 100.  

\begin{figure}
\begin{center}
  \epsfig{file=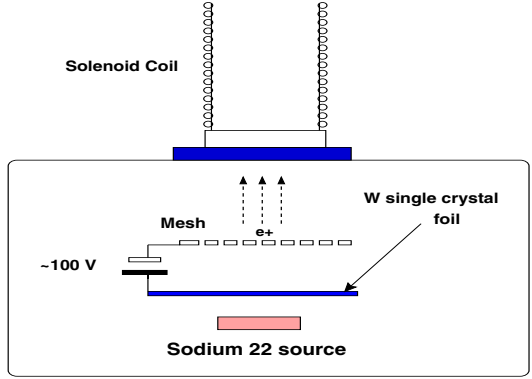,width=70mm,height=50mm}
\end{center}
\vspace{1.0cm}
 \caption{\em Schematic illustration of the W(100) single crystal  moderator.}
\label{moderator}
\end{figure}

The moderation and associated voltages applied as shown in 
Figure \ref{moderator} produce a positron beam which contains both
moderated positrons with energy of about 3 eV and an unmoderated ($\beta^+$
decay energy) positron component which has to be rejected.
There are several possibilities to build a positron velocity filter to select
the slow, moderated component and reject high energy component from the source.       We consider at the moment a simple bending filter, which is made of 
solenoidal coils positioned one after the other. Compensation of the 
beam drift due to the curved magnetic field is also foreseen.

The emittance of the beam is defined by the area in the phase space 
($r, \alpha$) multiplied by the square root of the energy and divided by a 
factor $\pi$, where $r$ and $\alpha$ are the radial positron position and 
angle of the beam.\ The brightness of the beam $B$ can be defined as
\cite{schultz} 
$B=\frac{I}{\phi^2 D^2 E}$,
where $I$ is the intensity, $E$ the particle energy, $\phi$ the angular 
divergence of the beam, and $D$ the beam diameter.\ The brightness of
the beam is limited by its initial emittance and by  
Liouville's theorem.
%which states that the density of points in the phase space remains constant 
%under the action of conservative forces (forces which can be derived from 
%a Hamiltonian) like electrostatic or magnetic forces.\ 
This means that if the 
beam size is decreased, the angular divergence is increased.\
%The initial emittance of the beam and characteristics of the moderated 
%positron trajectories. 
One of the problems is that positrons derived from the 
moderator will gain in emittance because of the presence of a  transverse 
electric field in the vicinity of the moderator mesh, see Figure 
\ref{moderator}.

%Another possibility is to build a  $E\times B$ velocity filter which 
% has been successfully built and 
%tested as described in \cite{filter}.\ 
%The device employs the usual crossed electric and 
%magnetic fields but, by incorporating a cylindrical electrode geometry, avoids
%the distortion and the consequent losses of intensity caused by post-filter 
%aperture that are common in conventional planar electrode 
%systems.

%% file: simulations.tex
\subsection{Analytical design of the pulsing system}

To avoid the already mentioned over-voltage problem with the post-amplifier, 
we try first  to design a simple pulsing system which  
 accelerates and/or decelerates positrons
 only in a single gap during 300 ns, such 
 that positrons arrive at the target after the buncher almost simultaneously.
This problem can be solved exactly if all positrons at
the entrance of the buncher have exactly the same
longitudinal momentum. We use a time dependent
electric field in the gap so that particles   arriving first are
decelerated and the later ones are accelerated. For a
given distance from the gap to the target  we have in this case one
free parameter - the deceleration potential at the time
$t=0$.  There is a
soft limitation in the choice of the initial value of the 
deceleration potential: the momentum of positrons coming first 
should not be very small, since a small
smearing of the initial momentum would cause  a large dispersion 
in time at the target.
%In our case particles coming to the buncher with $\simeq$100 eV 
%were initially decelerated by the 60 eV
%potential difference. 
However, we found 
the accelerating gap potential near  the end of the
300 ns time interval should be more than 1 kV, which is 
difficult to  achieve  in a high frequency pulsed mode operation.

  More convenient and economic is the buncher with two
modulation gaps. In this case, as is shown in Figure \ref{beam}, 
  the buncher consists
of the entrance drift tube 1, buncher electrode and exit drift tube 2.
The decelerating or accelerating potential is applied  to the buncher
electrode with respect to the drift tubes resulting in 
positron velocity modulation in two gaps.

The simulations of the extraction optics, beam transportation and  
 of the velocity modulation    
of positrons are performed with
 the GEANT4 \cite{geant} and 3D-Bfield \cite{mk} codes
with the goal to minimize the timing resolution and 
optimize the shape of the bunching pulse.

\begin{figure}
\begin{center}
  \epsfig{file=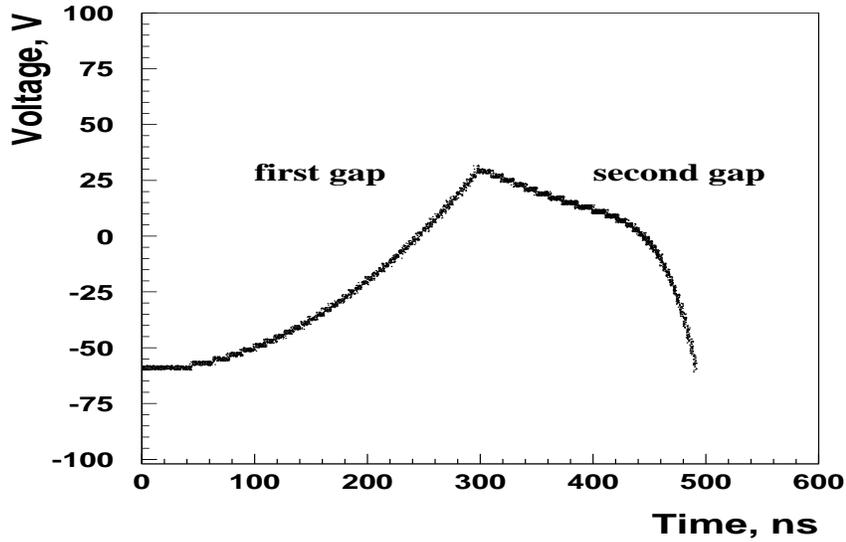,width=120mm,height=80mm}
\end{center}
\vspace{1.0cm}
 \caption{\em  The bunching voltages seen by positrons  at the 
first and the second velocity modulation gaps, respectively.}
\label{pulse}
\end{figure}

The numerical solution for the shape of the buncher pulse is shown in 
Figure \ref{pulse}. It was calculated for the  following characteristics 
of the system:
\begin{itemize}
\item moderated positrons have an average kinetic energy of $\simeq 3\pm0.5$ eV
and are emitted isotropically from the flat film of 
10 mm in  diameter. According to the previous measurements \cite{chen} the
low energy tail with intensity of 10\%  has been added,
\item the velocity filter tube axis is curved with
a radius of 45 cm.  Particles enter into the tube along Y (upward) and exit 
along the Z axis,
\item the buncher gaps are both 1 mm wide with the electric field
along the Z axis.
\item the  buncher electrode is 140 cm long,
\item the beam transportation tube is 1.05 m long and 10 cm in diameter
\end{itemize}

 The optimal duration of the buncher pulse and the shape of the potential
which positrons ''see'' in the  gaps are chosen to satisfy the following
 criteria:

\begin{itemize}
\item the time difference between two positrons arriving at the gaps 
should be smaller at the second gap for the applied potential,
\item the amplitude of the RF pulse should be within $\pm$60 V,
\item after modulation at the second gap, positrons should arrive 
 at the target at the same time.
\end{itemize}

 We chose the parabolic  time-dependent potential
proportional to $t^2$  and changing from from -60 V
(decelerating part) to 30 V  for the positron velocity
modulation at the first gap. Once this is fixed, 
the only free parameter left is the distance from the second gap to
the target. The time dependence of the potential at
the central electrode of the buncher at time $t>300$ ns can
then be calculated to make particles arriving
 at the target simultaneously. It is also assumed that
 the potential at the electrode at the
end of the bunching pulse  returns to its initial value
-60 V. The pulse shape at the second gap is found 
as the result of an iteration procedure for the solution of the 
 corresponding equations. Figure \ref{pulse} shows the resulting shape of the 
bunching voltages seen by positrons  at the 
first and the second velocity modulation gaps, respectively.

% In this case we apply time dependent potential to the
%central electrode of the buncher, other parts of the
%system being grounded. The gaps between the central
%electrode and other parts are the two
%decelerating/accelerating gaps of buncher. In this
%case there is much wider choice of solutions.

\begin{figure}
\begin{center}
  \epsfig{file=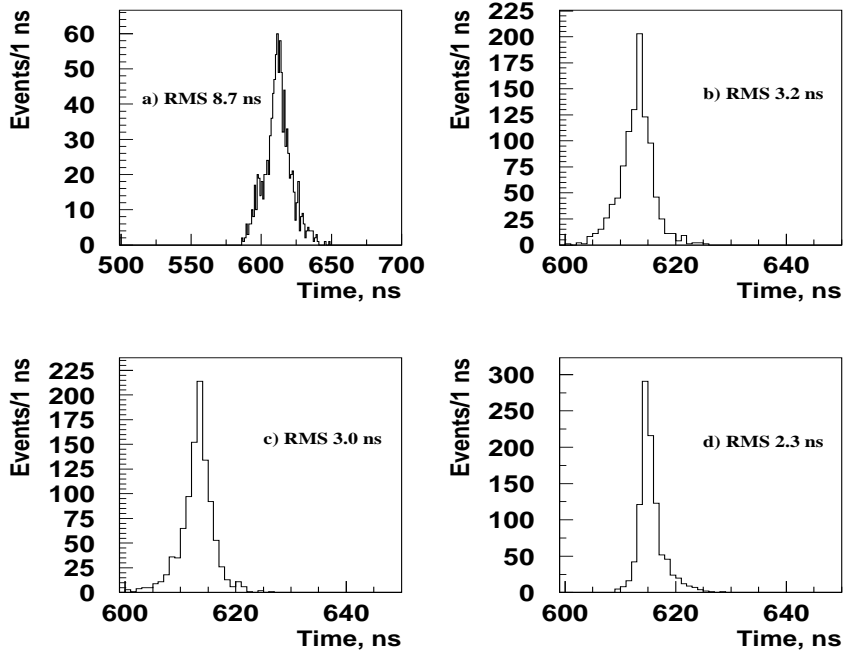,width=120mm,height=100mm}
\end{center}
\vspace{1.0cm}
 \caption{\em Distribution of the positron time-of-flight  
at the target for the following cuts on the longitudinal energy 
$E_{||}$ of positrons emitted from the moderator: 
a) $>0$ eV, b) $>$1.5 eV c) $>$2.0 eV and d)$>$ 2.5 eV. 
The initial positron energy distribution
 is Gaussian with an average of 3 eV and 
$\sigma = 0.5$ eV, the angular distribution is taken to be isotropic. }
\label{time}
\end{figure}

%% file: requirements.tex
\subsection{Requirements for the system components}

The time shape of the positron pulse at the target is affected by the
performance of the pulsed beam components.
 In this section requirements for the  monochromaticity of the
moderated positrons and for the amplifier used to generate the buncher
pulse  shown in Figure \ref{pulse} are considered.

\subsubsection{Monochromaticity of the moderated positrons}

%The slow positron beam is composed of a  $^{22}$Na positron source 
% and a single crystal W(100) foil of 1-3 $\mu$m thickness
%as shown in Figure \ref{moderator}.\ The foil is annealed at 2000$^o$C in a 
%vacuum of better than $\sim 10^{-8}$ Torr,and is attached in the front of the 
%sodium source to moderate positrons.\ A $\sim 10^{-4}$ 
%fraction of these energetic positrons emitted from the  $^{22}$Na source 
%is thermalized in the tungsten moderator and is re-emitted from the moderator 
%in a cone of about 30$^o$ as Monoenergetic, a  few eV slow-positrons.\\   

In Figure \ref{time} the simulated positron time distribution at the target 
is shown for different energy spread of the initial positrons on the moderator.  The degree of monochromaticity of the initial positrons was defined  by 
the cuts on the longitudinal positron kinetic energy $E_{||}$. The angular 
distribution of initial positrons was taken for simplicity to be isotropic. 
The results show that the best resolution achieved with 
this method is about 2 ns.   
The results also show that the quality of the moderator is an important 
parameter. So, the annealing of the W-foil {\em in situ} is 
important.\ Otherwise quick 
degradation of the surface quality through interactions with a gas 
results in a practically isotropic reemission of the positrons, and 
hence a significant increase in the phase space and timing spread.\
%The moderator heating could be performed  with a high-power 
%electron beam (50 W electron flashes) or  less preferable by heating 
%with a current placing the W moderator foil 
%between a folded 95\% transmission W-mesh. 
%%This mesh is  heated at 2000$^o$C for about 
%%2 min several times under high vacuum $\approx 10^{-9}$ Torr. 
% Laboratory  studies may be necessary to address some of these
%questions.

%Note, that recently the positron re-moderation efficiency
%of self-supported copper films $\sim 5000~ \AA$ thick was measured to be
%better than that of W \cite{brusa}.\ Taking into account that such  films
%are easily produced - for example their annealing  requires a relatively
%low temperature of $\simeq 450 ^{o}C$ - they are good candidates for
%positron beams with improved brightness.

\subsubsection{Buncher pulse and amplifier responce}

The post-amplifier choice is made on the basis of the following requirements.
We need an amplification factor of about 100 since the  typical amplitude 
of an AWG (arbitrary waveform generator)
is 1 V
and the output signal has a peak to peak value of 90 V. The gain has to be
programmable
for final tuning of the output amplitude. The amplifier has to be sufficiently 
powerful to drive the 50 Ohm load. A few tens of Watts
 is expected for a typical signal rate.
A wide frequency band from a few kHz to a few 100 MHz  is required to
minimize 
the signal shape distortions. And finally, the integral non-linearity
has to be within at most 1\% .

\begin{figure}
\begin{center}
  \epsfig{file=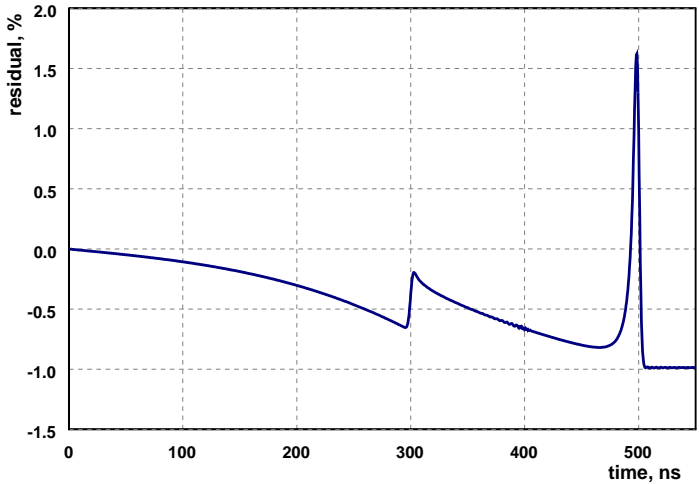,width=100mm,height=70mm}
\end{center}
\vspace{1.0cm}
 \caption{\em  Residual response  of the post amplifier 100A250A 
for the input pulse shown in Figure 3.}
\label{resid}
\end{figure}

We select commercial RF post-amplifier 100A250A  (Amplifier Research \cite{ra}) as a
 supplier of a fast pulse to the buncher. It has 100 W output
power and a 
frequency bandwidth from 10 kHz to 250 MHz. It is designed for 50 Ohm 
loading with reliable over-current and over-voltage protection. 
%The amplifier is controlled through either GPIB or RS232 lines.

The 100A250A post-amplifier is an AC-coupled device. To shift the signal
baseline to the -60 V level, an external DC source is used. 
This voltage supplier is
also used for fine pedestal tuning in order to compensate the baseline
shift in the case of a  high signal rate.

The response of 100A250A was simulated  following the circuit characteristics. 
The output pulse shape differs from the input due to the finite frequency
bandwidth. 
The residual shape is defined as $R(t) = S_{out}(t)-S_{in}(t)$
 where $S_{in}(t)$ is the input 
signal supplied by AWG with unit amplitude and 
$S_{out}(t)$ is the amplifier output pulse 
calculated for  unit gain. 
Figure \ref{resid} shows $R(t)$ expressed in percents of the input 
computed for the
signal shape shown in Figure \ref{pulse}.
 The input shape is parameterized by an analytical function
which has singularities (infinite first derivative) at $t =$ 300 ns
and $t = 500$ ns. 
Due to the upper frequency limit, the output signal is smoothed and differs
from the input  shape at these time points. 
This is indicated by two peaks on the residual
plot. The post-signal baseline shift of about 1\% is due to the lower frequency
limit.

One can see that the overall deviation of the response is not more
than about $\pm$1\%. It is expected that this value will be better for real
signals
because the 
AWG can not reproduce the singularities of the theoretical shape. The 
simulations of the beam showed that 1\% of signal deviation does not result
in
a significant distortion of the bunched positron pulse shape. The RMS of the 
corresponding distribution (see Figure \ref{time}, d) has been changed by 
less than 2\%. 
However, for a deviation of the order of 5\%, the RMS degrades from 2.3 to
2.8 ns. 
This means  that the shape of the buncher pulse must be reproduced
within about $\pm$1\% of the theoretical shape. 
This value seemes to be achievable \cite{ra}.

%% file: montecarlo.tex
\section{Monte Carlo simulations}

The positron  trajectories in the beam were simulated with 
the GEANT4 and the 3D-Bfield programs \cite{mk}. 
The $o-Ps$ production, propagation in the beam pipe, 
reflection on pipe walls and decay was simulated.
% with a special package. 
The events  for the $o-Ps\to 3 \gamma$ process were generated 
taking into account the decay matrix element and assuming decays at rest.

\begin{figure}[htb]
\begin{center}
\hspace{-0.cm}{\epsfig{file=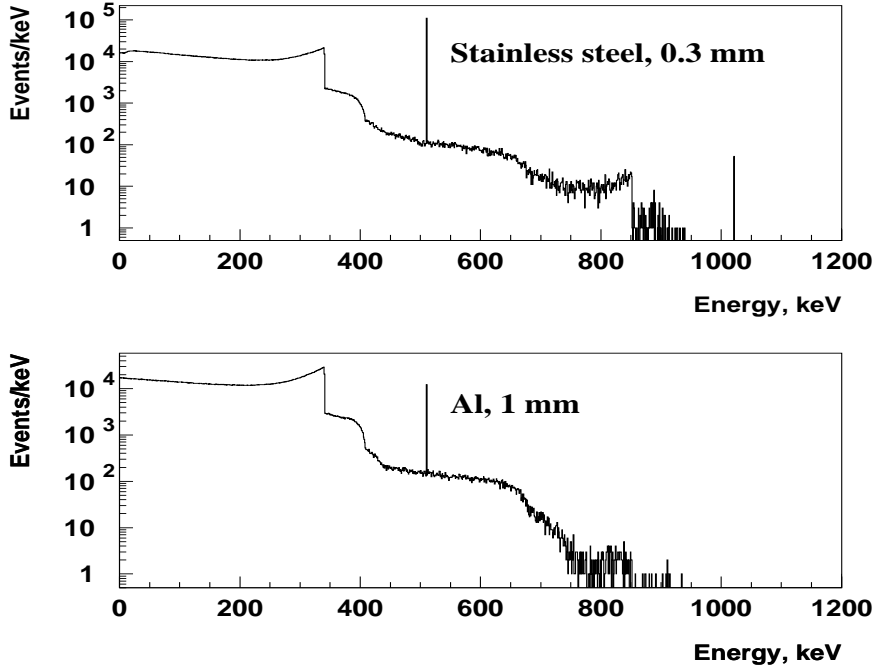,width=120mm,height=100mm}}
\end{center}
\vspace{1.0cm}
 \caption{\em Distributions of the energy deposited  in a 
0.3 mm thick stainless steel pipe (upper plot) and in a 
1 mm thick Al pipe. The total number of simulated $2\gamma$-events
is $10^8$ in both cases. The peaks at 511 keV and 1022 keV
correspond to the total photo-absorption either of a single 
511 keV photon or of both of them, respectively.}
\label{energy}
\end{figure}

The Monte Carlo simulation of the photon detection in the 
apparatus was based on 
the GEANT3  package \cite{geant}. The geometries of the beam transport
pipe,  photon detector, positron tagging system and its  material
were coded into simulations.
The simulation results were also benchmarked with the results of 
our previous experiment on a search for the o-Ps$\to \gamma+X_1+X_2$ decay mode
\cite{bader}. 
In Figure \ref{energy}, the distributions of the total energy deposited  in 
0.3 mm thick stainless steel pipe (upper plot) and in a 
1 mm thick Al pipe are shown for the 2$\gamma$ annihilation events. 
The  $\gamma$-detection inefficiency was found to be less than  
  $3 \times 10^{-8}$ for the case of an  Al 
 pipe assuming that the zero-energy signal is defined as an event with 
 the total 
energy deposition in the ECAL, $E\lesssim 100$ keV. 

 Simulations show that the 
main contribution to the $\gamma$-inefficiency comes from the total 
(due to photo-absorption)
  or fractional (due to Compton effect)   
photon energy loss in the material of the vacuum beam pipe.
In the latter case, the energy deposition in the ECAL corresponds to a 
small amount of photoelectrons, so that the ECAL signal could be comparable 
with ''zero-energy''  due to statistical fluctuations.
The $\gamma$-detection inefficiency  due to the limited ECAL thickness    
is estimated to be less than $10^{-8}$. 
A special mechanical design  discussed in section 3.2 is necessary to 
reduce absorption in the pipe by choosing the appropriate materials
and construction of the target region.
  
 The energy spectrum of  $o-Ps$'s produced at the MgO target is taken from 
 ref.\cite{mgo}. The angular distribution
is assumed to be isotropic and the intensity of the fast backscattered 
component \cite{vzg} is taken to be 1\% with a flat energy spectrum from 
10 to 100 eV for a primary positron energy of a few keV. 
The collisional $o-Ps$ dissociation probability for a $o-Ps$ kinetic 
energy greater than 10 eV is taken to be 100\%. 
The average number of $o-Ps$ collisions with the cavity walls 
during their lifetime in the cavity was estimated to be
 $N_{coll}\approx 2.5$, resulting in corresponding suppression 
factor in the sensitivity of the experiment (see sections 2,6). 
The other concerns are 
about  $o-Ps$ decays near the entrance aperture to the detector,
 i.e. in the region of
lower $\gamma$-detection efficiency,  
and the disappearence effect, namely when $o-Ps$'s escape the cavity 
region through the entrance aperture without being detected.

%% file: sensitivity.tex
\section{Sensitivity}

The experimental signature of  the $o-Ps\rightarrow invisible $ decay 
is an excess
 of events above the background at zero-energy deposition in the ECAL.
The 90\%-confidence level limit on the branching ratio for the 
$o-Ps \to invisible$ deacy for a background free  experiment is given by 
\begin{equation}
S(90\%)=\frac{N(o-Ps\to invisible)}{N_{o-Ps} N_{coll}}
\label{sens}
\end{equation}
where $N(o-Ps\to invisible)=2.3$ and 
the terms in the denominator are the integrated number of produced
$o-Ps$'s ($N_{o-Ps}$), and the average number of $o-Ps$ collisions in the
cavity, respectively. The number
$N_{o-Ps}$  is defined as a product 
$N_{o-Ps}=R_{e^+}\cdot \epsilon_{o-Ps}\cdot \epsilon_{e^+}\cdot t $, 
where the first factor is the  number of 
delivered positrons per second on the target, the second one is the 
efficiency for 
$o-Ps$ production, and the third one is the efficiency of the 
secondary electron
 transportation from the target to the MCP in the 
positron tagging 
system. Taking $R_{e^+}= 2\times 10^3/sec$, $\epsilon_{o-Ps}= 20\%$ and
$\epsilon_{e^+}= 100\%$, 
 we expect $\approx 7\times 10^7$ prompt and $\approx 1.7\times
 10^7~o-Ps$ annihilations per day. Thus, $S(90\%)\simeq 10^{-7}$.

 Eq.(\ref{sens}) gives the sensitivity for a background free
experiment. We expect
 backgrounds  which originates from the 
following sources:
i) a fake positron tagging ii) the annihilation energy loss
and iii) the disappearence effect. 
The are several sources of background that simulate the positron
appearance signal:
\begin{itemize}
\item the MCP after-pulses or pulses  
produced due to ionization of the residual gas by passing 
positrons;
\item cosmic rays;
\item environmental radioactivity.
\end{itemize}

The  MCP noise  is typically  low.
Its level depends, e.g. on the  $^{40}$K contamination in the 
surrounding materials and on the intensity of penetrating photons, although
the sensitivity of the MCP to photons is quite low.
However, for the purpose of this experiment the S/N ratio
must be better than $\simeq 10^7$. To reach this level a high 
SE emission coefficient of the target
 combined with a good MCP energy resolution
 and a high $o-Ps$ production rate is crucial. 
A preliminary estimate shows that taking into account the noise 
spectrum of MCP4655-12 and  the  average number of secondary 
electrons per positron for the MgO target to be $N_{SE}\simeq 4$ 
\cite{se},  the MCP S/N ratio
can be expected to be  better than $10^5$.
An additional noise suppression  factor of $\simeq$100 (at least) 
is expected from the use of the 
MCP signal in coincidence with the positron beam pulse  \cite{gninops} 
with the time resolution of $\simeq$ 3 ns.
The main source of  beam associated background is expected 
from  electrons and 
 ions due to ionization of the residual gas atoms by  
positrons. Thus, a good vacuum is important. 
We estimate that with $R_{e^+}\simeq 10^3/s$ positrons on target,
$\simeq 4$ secondaries per positron and a vacuum in the cavity of 
$\simeq10^{-8}$ Torr,  a MCP S/N ratio $>10^7$ is achievable.
The contribution from disappearence effect was also found to be small,
$\lesssim 5\times10^{-8}$, although not
completely free of assumption on fraction of fast $o-Ps$'s formed 
at the target.  
The preliminary overall background estimate results in 
$R_{bckg}\simeq 1.2~ events/day$. 
%Thus, the sensitivity at which one event of background is seen is estimated to be $S_B\simeq 10^{-7}$.

    In case of the observation of zero-energy events, one of the 
 approaches would be to measure their number  
as a function of the residual gas pressure in the cavity.\
This would allow a good cross-check:
relatively small variations of gas pressure results in 
 larger peak variations at zero energy due to the damping  of
 $o-Ps\to o-Ps'$ oscillations.

%% file: conclusion.tex
\section{Summary}

The design of our experiment to search for the $o-Ps\to invisible$ decay mode 
in vacuum has been  presented. The  sensitivity of the experiment 
to the branching ratio of the $o-Ps \to invisible$ decay in vacuum around
 $\simeq 10^{-7}$ seems to be achievable. Assuming that 
$o-Ps\to o-Ps'$ oscillations occur with  mixing strength value 
$\epsilon \simeq 4\times 10^{-9}$ a total number of $\simeq$ 100 signal 
events would be expected in the ECAL during one month of data taking. 
Given that about  40 background 
events are expected a discovery with  significance $\simeq 11$
  could be  possible \cite{bk}. 
% In case of no signal observation, the obtained upper limit will
%constrain the existence of a mirror sector outside the standard model
%supporting the idea of an asymmetric vacuum. 

One of the main features of the experiment is the use of a 
high-efficiency pulsed positron beam. 
The new proposed pulsing method allows 
to compress a 300 ns initial positron pulse into a pulse with 
$<$ 3 ns width. This will allow  to enhance the signal-to-noise ratio
for the efficient tagging of $o-Ps$ production by more than an 
order of magnitude. A modification of the pulsing system 
could be made to use the beam for applications for  material studies. 
    Different components of the detector have to be 
constructed and fully simulated, in order  
 to test several crucial points of the design.